\def\k{\vec{k}\/}

\def\be{\begin{equation} }
\def\ee{\end{equation} }
\def\eq{\ =\ }
\def\n{\noindent }

\def\JPC{{\it J. Phys. C : Solid State Phys\ }\/}
\def\JPF{{\it J. Phys. F : Metal Phys\ }\/}
\def\JPCM{{\it J. Phys. Condens Matter\ }\/}
\def\PR{{\it Phys. Rev.\ }\/}
\def\PRL{{\it Phys. Rev. Lett.\ }\/}

\documentclass{elsart}
\usepackage{epsfig}
\usepackage{graphicx}
\usepackage{amssymb}

\begin{document}
\begin{frontmatter}

\title{Study of phase stability of MnCr using the augmented space recursion based orbital peeling technique} 
\author[label4]{Rudra Banerjee}
\author[label4,label1]{Abhijit Mookerjee}
\address[label1]{Department of Materials Science,  S N Bose National Centre for Basic Sciences,Block JD, Sector III, Salt Lake, Kolkata 700098,India}
\address[label4]{Advanced Materials Research Unit, S N Bose National Centre for Basic Sciences,Block JD, Sector III, Salt Lake, Kolkata 700098, India}

\begin{abstract}
In an earlier communication we have developed a recursion based approach to the study of phase stability and transition
of binary alloys \cite{fecr}. We had combined the recursion method introduced by Haydock, Heine and Kelly\cite{hhk} and the
our augmented space approach\cite{asr} with the orbital peeling technique proposed by Burke \cite{burke} 
to determine the small energy differences involved in the discussion of phase stability. 
We extend that methodology for the study of MnCr alloys.
\end{abstract}
\end{frontmatter}

\textwidth 20cm
\baselineskip 12pt

\section{Introduction} 
In an earlier communication \cite{fecr} we had introduced 
 the augmented space recursion (ASR) \cite{asr} based orbital peeling method (OP) \cite{burke} as an useful and numerically accurate method for the  
calculation of the `pair energies'. This  allowed us to map the binary alloy problem onto an 
effective Ising-like model and study the stability or otherwise of different ordered phases that might 
arise if the disordered alloy is cooled below  some critical temperature.  The aim of this paper is to extend the use of this
method to study MnCr alloys.

 These pair
energies are small differences of relatively large electronic energies and a brute force calculation
is likely to yield these small numbers with errors which are usually of the order of or larger than the numbers 
themselves. Direct estimations like the OP method are therefore appropriate in these situations.
Our earlier analysis of FeCr, FePd and PdV alloys illustrated the success of the  ASR-OP method in 
predicting the low temperature phases of these binary alloys and their stability.
The analysis  involved ordering and mixing energies  obtained from a generalized perturbation framework (GPM) \cite{duc,turchi} and the Fourier transform $V(\k)$ of the pair energies which are related to diffuse scattering
 intensities as described by Krivoglatz, Clapp  and Moss (KCM) \cite{cm}. The basic ideas behind these 
methods  have been described in detail in the references quoted above. 
For the sake of completeness, we shall briefly dwell upon the main points used in our analysis. 

\begin{figure}
\centering
\resizebox{4in}{3.5in}{\includegraphics{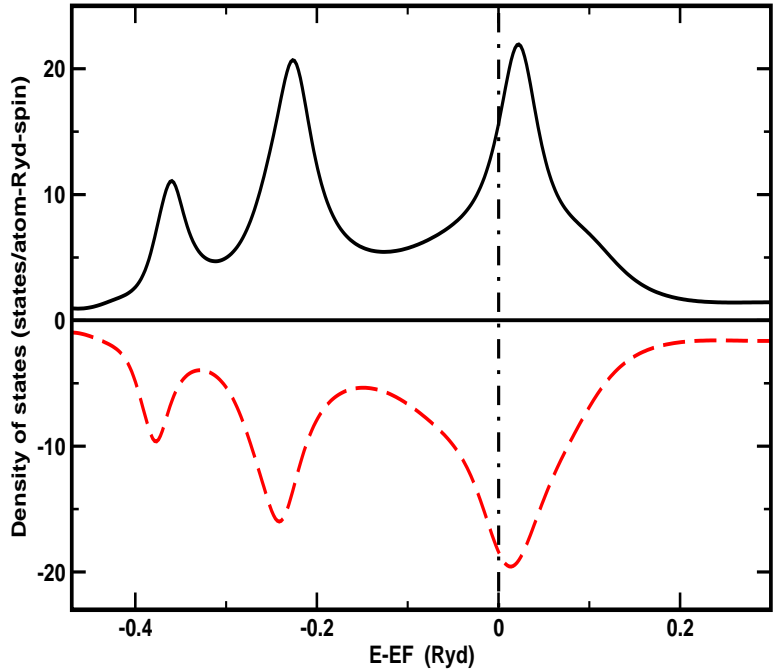}}
\caption{\label{fig1}(Color Online)  Densities of states for a series of alloy compositions
for 50-50 MnCr alloy.}
\end{figure}

The basis of our subsequent analysis is the electronic structure calculation on the MnCr alloy system. We
have chosen the tight-binding linear muffin-tin orbitals (TB-LMTO) approach \cite{tblmto}  which 
provides a first-principles, density functional based  
tight-binding sparse representation of the Hamiltonian. Such a Hamiltonian is appropriate for
describing the electronic structure of random substitutional 
 alloys in which the configurational fluctuations due to disorder are {\sl local}. 
The sparseness of the TB-LMTO Hamiltonian  is a suitable  input for the recursion method.

Realistic models for disordered alloys require us to go beyond the single-site mean-field approximations and treat effects of configuration fluctuations more accurately. This we shall attempt through the ASR \cite{tf, sasnet}. As mentioned earlier, the pair-energies have to be  accurately determined. We shall adopt the recursion
based OP for such calculations.
 The approach followed here will provide
 a unified recursion based methodology to address such problems.

In this communication our choice of system is the MnCr alloy. We shall first obtain the configuration averaged density of states for a series of alloy compositions using the TB-LMTO-ASR. The ASR has been discussed in great detail earlier. We shall only stress here that it generalizes the single-site mean-field approximations to include the effect of the configuration fluctuations of the
near neighbourhood of a site and yields configuration averaged Green functions which retain their analytic and lattice translational symmetries even after approximation. The results for the 50-50 alloy are shown in
Fig.  \ref{fig1}. The converged potential parameters are input into the OP calculations that follow.

We shall organize the paper as follows : in section 2 we shall describe the OP-ASR method for obtaining the effective pair energies, the expression for the ordering and mixing energies and
the analysis  for the Fourier transform of the pair energies. In the section 3 we shall analyze
 the results for the equi-atomic MnCr alloy.  Concluding remarks will be given in the final section.

\section{Methodology}

\subsection{Total energy and pair energies}

The simplest model which analyzes the emergence of long ranged order from a disordered
phase is the Ising model. Our approach will attempt  to map the energetics of the binary alloy
problem onto an equivalent `spin-half' Ising model.
We need a derivation of  the  lowest  configurational
energy  for the  alloy system in terms  of  effective
multi-site   interactions,   in   particular   ``effective pair 
energies" (EPE) \cite{epi}. We need to   accurately 
and reliably  determine   the EPE.
Our approach will be  to start  with  the
disordered  phase,  set  up  a  perturbation  in   the   form   of
concentration fluctuations associated with an ordered  phase  and, from its
response, 
study whether the alloy can sustain  such  a  perturbation.   This
approach includes the generalized perturbation  method (GPM) \cite{kn:gpm},  the
embedded cluster method (ECM) \cite{kn:ecm} and the concentration  wave  approach
\cite{kn:cwm}. 

 We shall begin with a homogeneously disordered alloy A$_x$B$_{1-x}$, where every site is occupied by either an  A or a B type of atom with probabilities proportional to their concentrations. We define the `occupation' variable $n_{\vec{R}}$ to be a random variable which takes on the values 1 and 0 according to
whether the site labeled $\vec{R}$ is occupied by an A or a B atom. Its average
$\ll\!\! n_{\vec{R}}\!\!\gg = x$. This  perturbative approach expands the total internal energy of a particular atomic configuration as follows :

\be
E\eq V^{(0)} + \sum_{\vec{R}} V^{(1)}_{\vec{R}} \ \delta n_{\vec{R}} + \frac{1}{2}\sum_{\vec{R}}\sum_{\vec{R'} \ne \vec{R}} V^{(2)}_{\vec{R}\vec{R'}} \ \delta n_{\vec{R}}\ \delta n_{R'} + \ldots
\label{pair}
\ee

\n here $\delta n_{\vec{R}} = n_{\vec{R}} - x$ and $\ll\!\! \delta n_{\vec{R}}\!\!\gg = 0$.
If the configuration is homogeneously disordered then it immediately follows that  $\ll\!\! E\!\!\gg = E_{dis} = V^{(0)}$.
From the above definition we can interpret the other two expansion terms as follows :  if $E^I$ is the configuration averaged total energy of a configuration in which any arbitrary site labeled $\vec{R}$ is
occupied by a atom of the type $I$ and the other sites are randomly occupied, and
$E^{IJ}$ is the averaged total energy of another configuration in which the sites $\vec{R}$ and $\vec{R'}$ are occupied by atoms of the types I and J respectively and all other sites
are randomly occupied, then from equation (\ref{pair}) it follows that :

\be V^{(1)}_{\vec{R}} \eq E^A - E^B \qquad V^{(2)}_{\vec{R}\vec{R'}} \eq E^{BB}+E^{AA}-E^{AB}-E^{BA}\label{eq2}\ee

The one-site energy $V^{(1)}_{\vec{R}}$ is unimportant for bulk ordered structures emerging from disorder. It is
important for emergence of inhomogeneous disorder at surfaces and interfaces \cite{indra}. The pair energies
$V^{(2)}_{\vec{R}\vec{R'}}$ are the most important factors governing emergence of bulk ordering.

The interpretation of equation (\ref{eq2}) immediately allows us to introduce a method to obtain the
pair potentials directly rather than calculate the total energies and then subtract them. Since they are small differences (of the order of mRyd) of large energies (of the order of $10^3$ Ryd), a direct calculation will produce errors larger than the differences themselves. The orbital peeling method (OP)\cite{burke} based on recursion \cite{hhk} was introduced by Burke precisely to calculate such small differences, albeit in a different situation.

The total energy of a solid may be separated into two terms : a
one-electron band contribution E$_{BS}$ and an electrostatic
term E$_{ES}$ which includes several contributions : the Coulomb
repulsion of the ion cores, the correction for double counting
terms due to electron-electron interaction in E$_{BS}$ and a
Madelung energy in case the model of the alloy has atomic spheres which are not
charge neutral. The
renormalized cluster interactions defined in equation  (\ref{pair})
should, in principle, include both E$_{BS}$ and E$_{ES}$
contributions. Since the renormalized cluster interactions
involve the difference of cluster energies, it is usually assumed
that the electrostatic terms cancel out and only the band
 contribution is important. Obviously, such an
assumption is not rigorously true, but it has been shown to be
approximately valid in a number of alloy systems \cite{turchi}. 
We shall accept such an assumption and  our stability  arguments 
starting  from the disordered side, will be  based on the band
structure contribution alone. 

 The effective  pair  interactions  can  be  related  to
the change in the configuration averaged local density of states :

\be V_{\vec{R}\vec{R'}}^{(2)}\enskip = \enskip   \int_{-\infty}^{E_{F}} dE\ (E-E_{F})\ \Delta n(E)
\label{eq3}\ee

where $\Delta n(E)$ is  given by :

\[
\Delta n(E) = -\frac{1}{\pi}\ \Im m\ \sum_{IJ}^{AB}  \mbox{Tr} %
\ll\!\! (EI-H^{(IJ)})^{-1}\!\!\gg \xi_{IJ}
\]

$\xi_{IJ}= 2\delta_{IJ}-1$, i.e. is $\pm$ 1 according to whether $I=J$ or $I\not= J$.
 There are four possible pairs  $IJ$~: AA, AB, BA and BB. H$^{(IJ)}$
is the Hamiltonian of a system where all sites except $\vec{R}$ and $\vec{R'}$ are
randomly occupied. The sites labeled $\vec{R}$ and $\vec{R'}$  are occupied by
atoms of the type $I$ and $J$. This change in the averaged local density of
states can be related to the generalized phase shift $\eta$(E) through the
equation :

\[
\Delta n(E) \eq {{d\eta (E)} \over {dE}} = \frac{d}{dE}\ \left\{\log {{\det \ll G^{AA}(E)\gg \det\ll G^{BB}(E)\gg} \over {\det
\ll G^{AB}(E)\gg \det\ll G^{BA}(E)\gg}}\right\}
\]

G$^{IJ}(E)$ is the resolvent of the Hamiltonian H$^{(IJ)}$. The  generalized  phase  shift  $\eta(E)$  can be  calculated
following the orbital peeling method of  Burke \cite{burke}. 
We  shall quote only the final result : The pair energy function is defined as :

\begin{eqnarray}
 f_{\vec{R}\vec{R'}}(E) & = & f(\vec{R}-\vec{R'}) = \sum_{IJ}^{AB} \sum_{\alpha = 1}^{L_{\rm max}}
\xi_{IJ} \int_{-\infty}^{E}dE'\ (E'-E) \log \ll G_{\alpha}^{IJ}(E')\gg\nonumber\\
 & = &  \sum_{IJ} \sum_{\alpha = 1}^{L_{\rm max}}
                 \left[ \sum_{k=1}^{p-1} Z^{\alpha,IJ}_{k} - \sum_{k=1}^{p}
                 P^{\alpha,IJ}_{k} + \left( N^{\alpha,IJ}_{P} -
                 N^{\alpha,IJ}_{Z} \right) E \right]
\end{eqnarray}

 $\ll G_{\alpha}^{IJ}\gg (E)$ denote  the  configuration averaged resolvents in
which the orbitals from $L$ = 1 to $(\alpha-1)$ are deleted.
$Z^{\alpha,IJ}_{k}$ and $P^{\alpha,IJ}_{k}$ are its zeros and
poles and 
$N^{\alpha,IJ}_{Z}$ and $N^{\alpha,IJ}_{P}$ are the number of
such zeros and poles of $\ll G_{\alpha}^{IJ}(z)\gg $ below $E$. The zeros and poles are
obtained directly from the recursion coefficients for the averaged resolvents and these
are obtained from the TB-LMTO-ASR.
  This method of zeros and poles  enables one  to
carry out the integration in equation (\ref{eq3}) easily avoiding the multi-valuedness  of
the integrand involved in  the  evaluation  of  the  integral  by
parts.  

The pair energy is then given by : $V_{\vec{R}\vec{R'}}^{(2)} = f_{\vec{R}\vec{R'}}(E_F) =
V^{(2)}(\vec{R}-\vec{R'})$. The last expression follows if the background system, in which
the A and B type atoms are immersed at $\vec{R}$ and $\vec{R'}$, is homogeneously
disordered.

\subsection{The Krivoglatz-Clapp-Moss analysis}

Philhours and Hall \cite{ph} have suggested, and Clapp and Moss \cite{cm} have formally shown  that a sufficient (but not necessary) condition for a stable ground state is that the
wave vector of concentration waves corresponding to an ordered phase lie in the positions of the minima of the Fourier transform of the pair energy function

\[ V(\vec{k}) = \sum_{\vec{R}-\vec{R'}} \exp\left\{i\vec{k}\cdot(\vec{R}-\vec{R'})\right\}\  V^{(2)}(\vec{R}-\vec{R'})\]

The above  statement follows from the expression for the inverse susceptibility which measures the
response of the disordered system to the concentration fluctuation perturbation described above.

\[
 \chi^{-1}(\vec{k}) \propto 1+x(1-x)\beta V_{\rm eff}(\vec{k})
\]

In a zero-th approximation  $V_{\rm eff}(\k) = V(\k)$. Corrections to the effective
 pair-function has been described in detail by Chepulskii and Bugaev \cite{cb}. 
We have used here the Ring Approximation suggested by the authors as the one most
suitable for our analysis :

\begin{equation}
V_{\rm eff}(\vec{k}) =V(\vec{k}) - (\beta/2)(1-2x)^2 \ \int\frac{d^3\vec{q}}{8\pi^3}\ F(\vec{q})F(\vec{k}-\vec{q})
\end{equation}
\n where 
\[ F(\vec{q}) = \frac{V(\vec{q})}{1+x(1-x)\beta V(\vec{q})}
\]

\subsection{Ordering and mixing energies}

Finally, the GPM expression also gives the ordering  energy~:

\[ \Delta E_{ord} = \frac{1}{2} \sum_n V_{0n}^{(2)} Q_n \]

where $n$ is a $n$-th nearest neighbour of an arbitrarily chosen site (which we label 0) and
$Q_n = (x/2)(N_n^{BB} - xN_n)$, $N_n^{BB}$ is the number of BB pairs and
 $N_n$ the total number of pairs in the 
$n$-th nearest neighbour shell of 0.  

With reference to the total energies of the pure constituents, in the approximation where we
only restrict ourselves to pair energies and ignore all three body energies and higher, the so-called mixing energy is given by : 

\[ \Delta E_{mix} = - \frac{1}{2} x(1-x) \sum_n\ N_n V_{0n}^{(2)} \]

The averaging procedure ASR has been described in great detail in many earlier papers and we refer the reader to
the review \cite{tf} in which the method and its relation to the CPA and its generalizations has been discussed
extensively.

\section{Results and Discussion}

\begin{figure}[t]
\centering
\vskip 1cm
\resizebox{3in}{2.5in}{\includegraphics{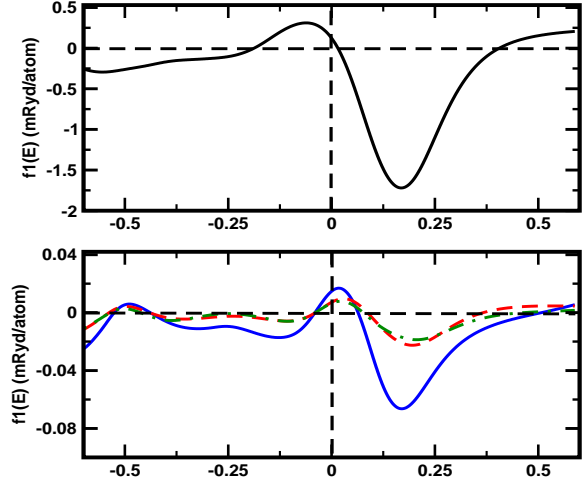}}
\caption{\label{fig2}(Color Online)  Pair functions calculated for Mn$_{50}$Cr$_{50}$, variation shown against E-E$_F$, calculated from TB-LMTO-ASR-OP.
(top) Nearest neighbour pair function at a distance $\sqrt{3}a/2$, $a$ being the equilibrium lattice parameter (bottom) second, third and fourth nearest neighbour pair functions at distances $a$, $\sqrt{2}a$ and $\sqrt{3}a$}
\end{figure}

 We have calculated the composition dependent pair energy functions using the TB-LMTO-ASR coupled
with the orbital peeling technique and the result for the equi-atomic composition is  shown in Fig. \ref{fig2}
The nearest neighbour pair energy function $f_1(\vec{R}-\vec{R'},E)$ shows the characteristic shape of a positive lobe, indicating
ordering, near the position of half filling, flanked by negative lobes indicating segregation near
empty and complete filling fractions.

\begin{figure}[t]
\centering
\vskip 1cm
\resizebox{3in}{2in}{\includegraphics{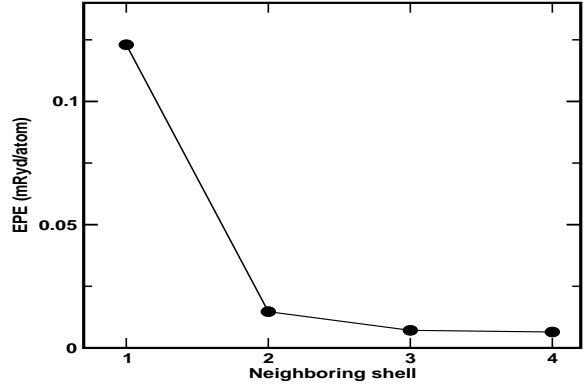}}
\caption{\label{fig3}  Pair energies  calculated for MnCr at a 50-50 composition.}
\end{figure}

We should note that in our approach, both the pair energy function itself and the position of the Fermi energy depend upon the composition of the alloy and its band filling.  
 This is in contrast to some analysis (like the Connolly-Williams) which depend on similar, but composition independent, pair energy functions. The
Fig. \ref{fig3} shows the pair energies $V^{(2)}(\vert \vec{R}-\vec{R'}\vert)$ for the equi-atomic composition. 

\begin{table}
\centering 
\caption{\label{tab1}Weights for different neighbouring shells for seven different bcc based equi-atomic superstructures.}
\begin{tabular}{ccccc}\hline\hline
         & \multicolumn{4}{c}{Neighbouring shells} \\ \hline
Structure \phantom{X}&\phantom{X} 1 \phantom{X}&\phantom{X} 2 \phantom{X}&\phantom{X} 3 \phantom{X}&\phantom{X} 4 \phantom{X} \\ 
\hline
  Segregated &   1.00 &  0.75  & 1.50 &  3.00   \\
  B2 &            -1.00 &    0.75 &   1.50&    -3.00 \\
  B32 &           0.00  & -0.75  &  1.50 &   0.00   \\
  B11 &           0.00  &  0.25  & -0.50  &  0.00   \\
  ST1 &           -0.50  &  0.25  &  0..00 &  0.50   \\
  ST2 &           0.00  & -0.25  & -0.50 &  0.00   \\
  ST3 &           0.50 &  0.25 &  -0.00 & -0.50  \\ \hline
\phantom{X}&&&& \\
\end{tabular}
\end{table}

\begin{figure}[b!]
\centering
\vskip 2cm
\rotatebox{0}{\resizebox{3.5in}{2.5in}{\includegraphics{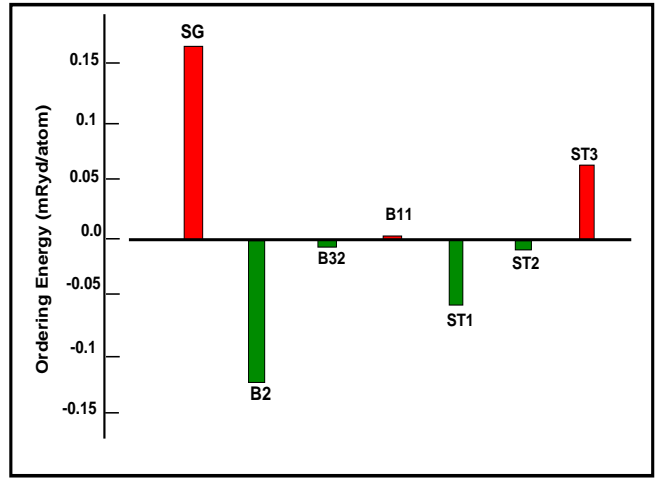}}}
\caption{(Color Online) \label{str} Ordering energies for seven different structures for MnCr based on Table 1.}
\end{figure}

The behaviour of the pair energies for MnCr indicate ordering tendencies up to the fourth nearest
neighbours.
The pair energies rapidly converge to zero with distance. In fact, although
we had calculated the pair energies up to the seventh nearest neighbour shells, their values beyond the fourth shell were smaller than the error bars of our
calculational method and therefore these numbers were not really reliable and were not used for our analysis. The same is true for the results of the ordering energies for seven different
structures and super-structures based on the body-centered cubic lattice. We have, therefore, calculated
the ordering energies with contributions only up to the fourth neighbouring shell.  The Table \ref{tab1} gives us the weights $Q_n$ for the seven bcc based structures
 required to obtain the ordering energies in this alloy system. These superstructures are described in detail by Finel and Ducastelle \cite{fd}.

Based on the Table \ref{tab1} we present the ordering energies for the same seven bcc based structures for MnCr in Fig. \ref{str}. 
Unlike our earlier study of FeCr which indicated segregation and possible ordering in the ST3 superstructure, for MnCr ordering
in the B2 structure with a possible competition from the superstructure ST1 is the most energetically probable. The contrast between
the two alloys FeCr and MnCr is interesting : Fe segregates with Cr while Mn orders. The stainless steel alloys, a class of 
which are ternary alloys FeCrMn should then see competition between segregation and ordering. We intend to study the ternary compositions
subsequently.

\begin{figure}
\centering
\rotatebox{270}{\resizebox{2.5in}{2.5in}{\includegraphics{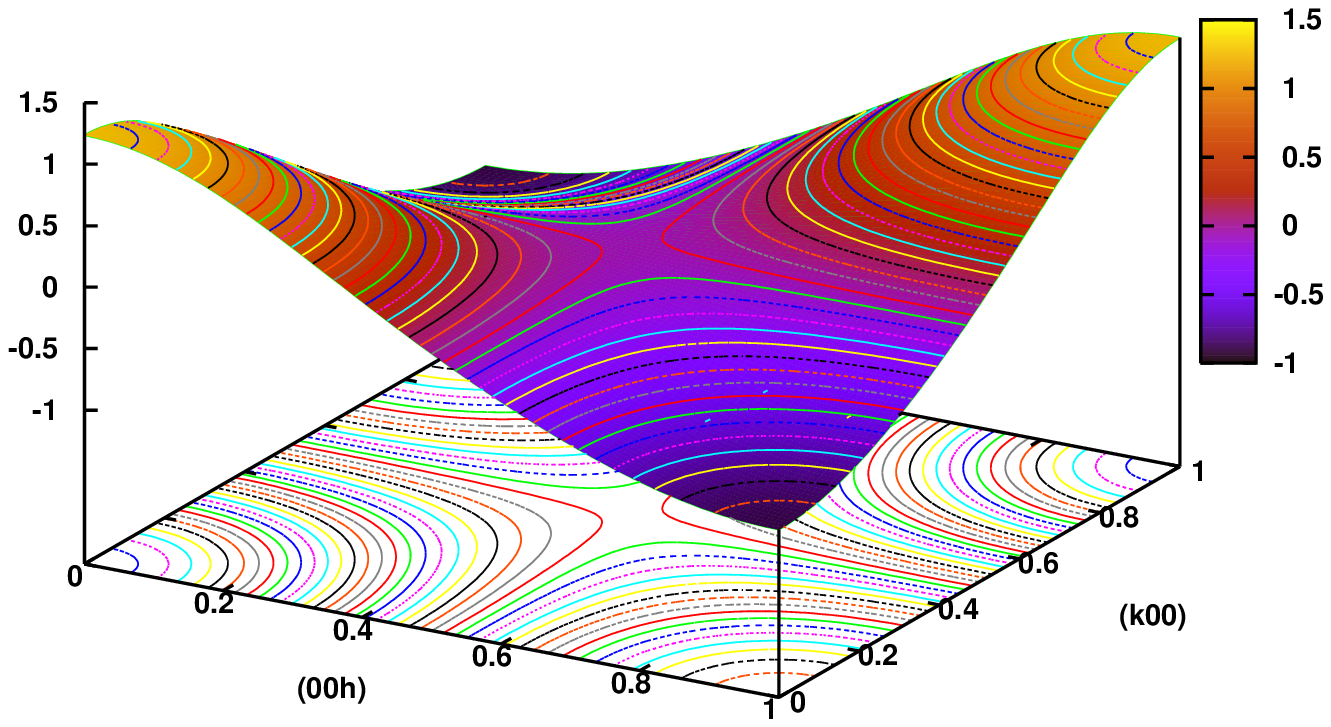}}}
\rotatebox{270}{\resizebox{2.5in}{2.5in}{\includegraphics{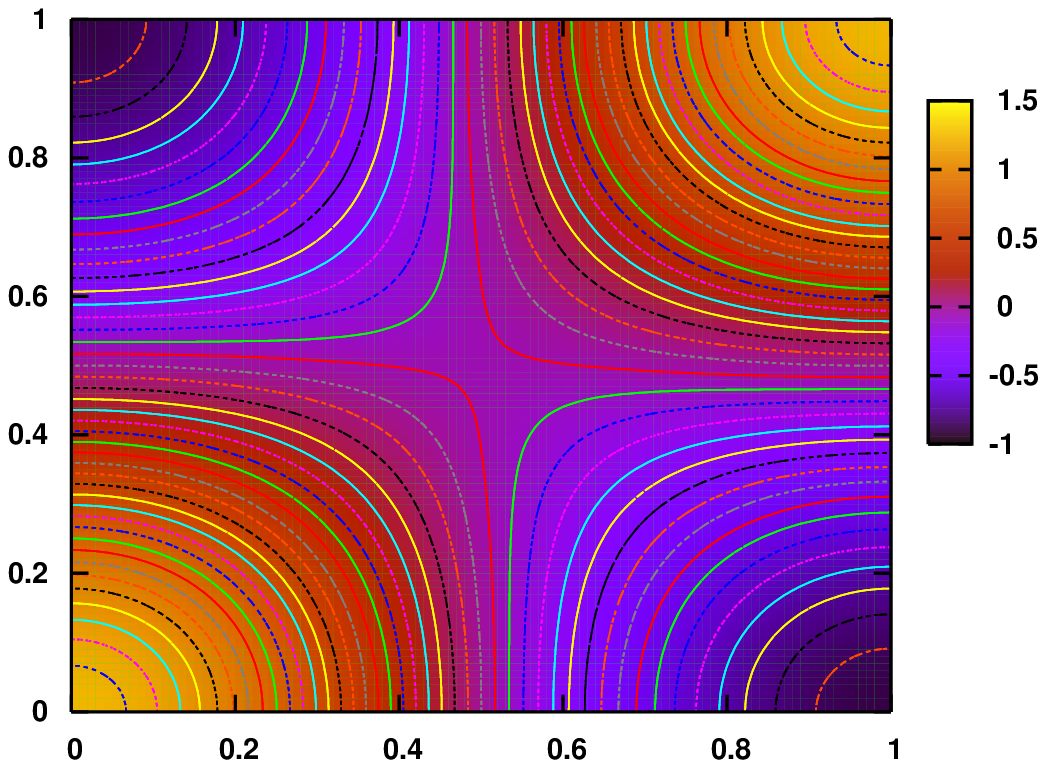}}}\\
\rotatebox{270}{\resizebox{2.5in}{2.5in}{\includegraphics{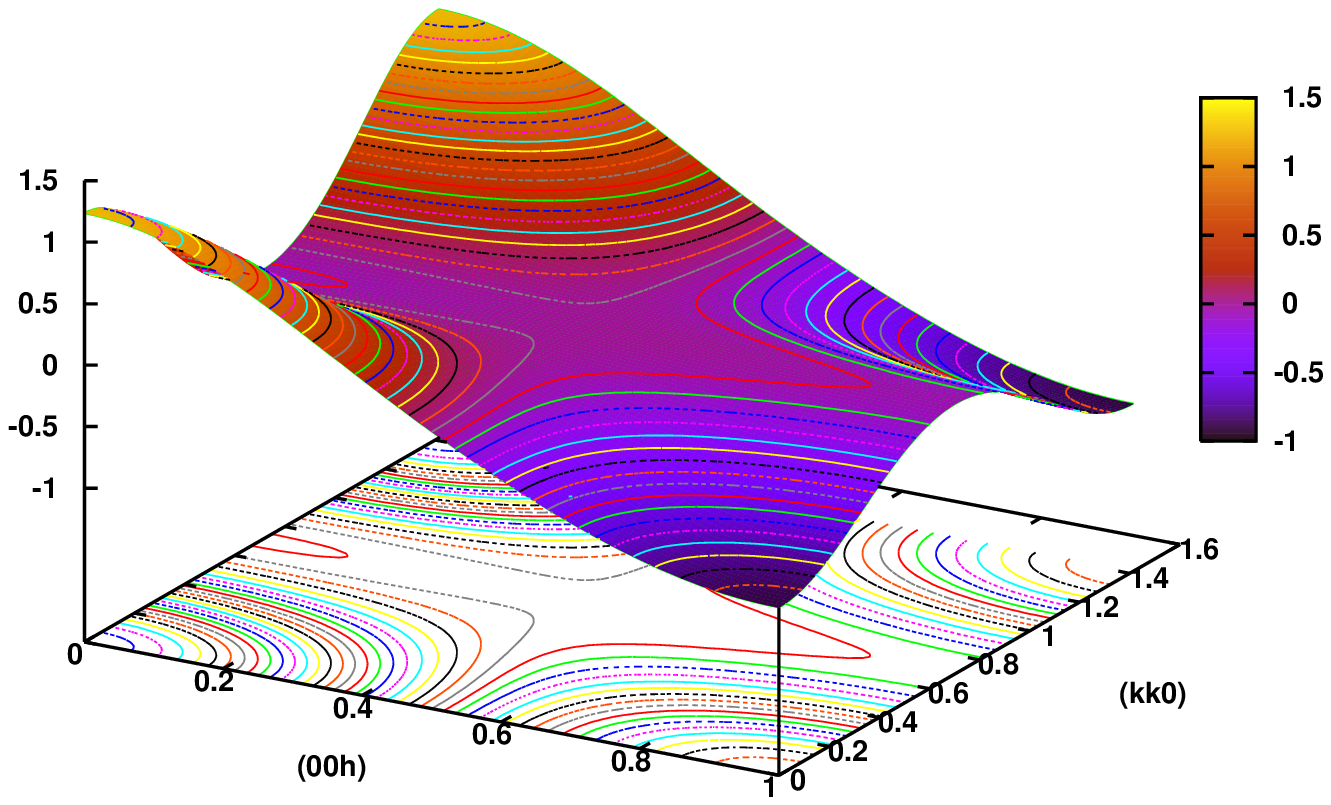}}}
\rotatebox{270}{\resizebox{2.5in}{2.5in}{\includegraphics{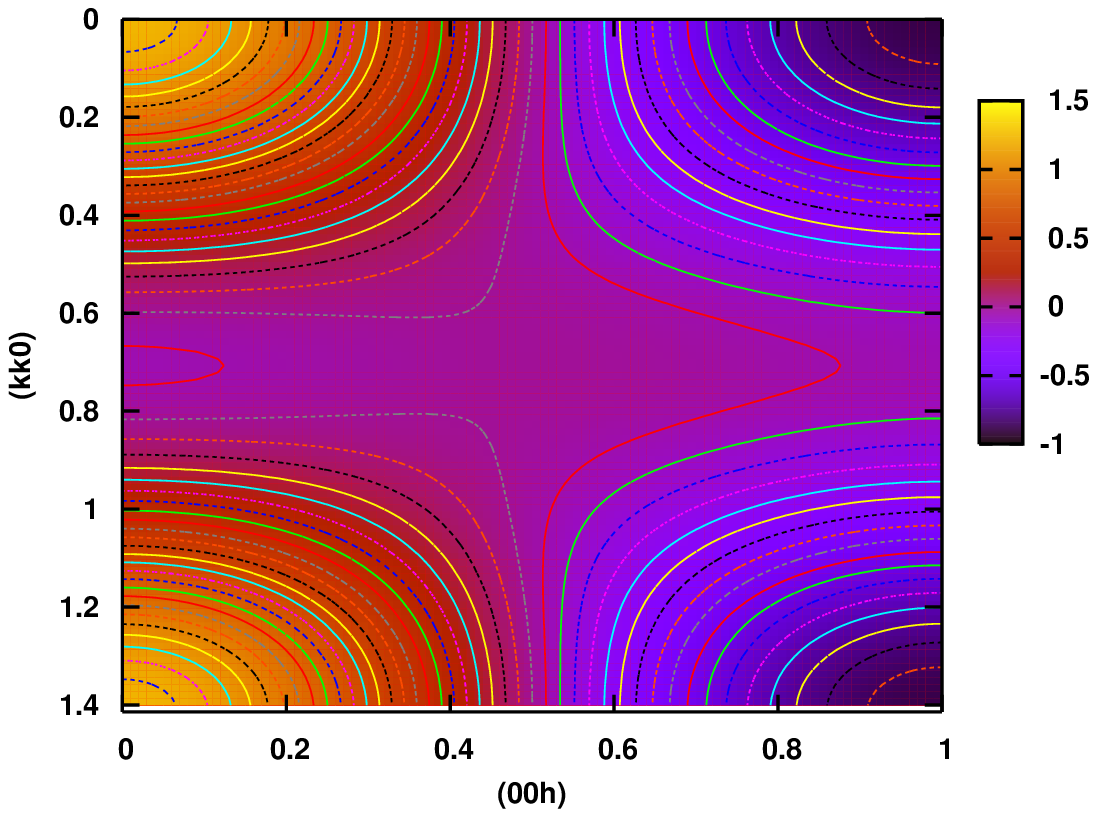}}}
\caption{(Color Online) \label{fig4} $V_{\rm eff}(\vec{k})$ and contours on the plane bounded by (top) (001) and (100)
and (bottom) (001) and (110)}
\end{figure}

Finally we shall examine the Fourier transform $V(\vec{k})$ of the pair energies and carry out a Clapp-Moss type of analysis. The contour diagrams 
for Fourier transform $V(\vec{k})$ in the 
is plotted  in Fig. \ref{fig4} for the (001) and  (1$\bar{1}$0) planes. The minimum occurs at $\vec{k} = (001)$ This is indicative
of a possible B2 type of ordering, as indicated by the ordering energy analysis.

In order to ascertain whether the ordered state associated with the minimum is stable compared to the segregated species, we have to carry out
a much more detailed analysis including the contribution of the energy of mixing. 
The mixing energy for the B2 structure  is 0.16412 mRyd/atom
 This indicates that ordering is stable against segregation.

We should comment here that the above discussion is based only on the electronic contribution
to the diffuse scattering intensity. At the temperatures
that we are interested in there are contributions from the vibrational excitations in the system.
 Our aim here was to indicate the possibility of ordering tendencies
in MnCr rather than accurate estimation of energetics and transition temperatures.  In any
detailed and accurate statistical mechanical calculations we must include the contribution of vibrational
excitations to the free energy of the alloy.

\section{Conclusion}
In a previous work \cite{fecr} we had introduced and examined the suitability and accuracy of the 
Augmented Space Recursion (ASR) based Orbital Peeling (OP) method for the generation of pair energies.
In this communication we have extended these ideas into body-centered cubic MnCr alloy.
We have looked at the phase stability of the equi-atomic alloy. Unlike our previous work on
FeCr, FePd and PtV alloys,  for MnCr we had no earlier theoretical work to compare with.
However, since our applications to FeCr, FePd and PtV gave us
satisfactory results, in good agreement with   
experimental evidence, we have confidence in our present results. With FeCr, this work will
form the background of our extension of this work to ternary FeMnCr stainless steel alloys.
 The recursion and ASR is now available with both relativistic
corrections (including spin-orbit terms)\cite{huda} and non-collinear magnetism \cite{bergman,
tarafder}. Since earlier we have proposed the ASR as a 
analyticity preserving generalization of the single-site mean-field coherent potential
approaches, this work will provide further incentive to extend the use of the ASR to problems
beyond the simple density of states calculations and in problems were relativistic corrections
and non-collinear magnetism will play significant roles.

\end{document}